\documentclass[11pt]{article}
\usepackage{hyperref}
\pdfoutput=1
\begin{document}
\title{Do Proximate Micro-Swimmers Synchronize their Gait?}
\author{Jinzhou Yuan$^1$, Kun He Lee$^2$, \\ David M. Raizen$^3$ and Haim H. Bau$^1$ \\
\\\vspace{6pt} $^1$Mechanical Engineering and Applied Mechanics, \\ $^2$Department of Bioengineering, \\$^3$Department of Neurology, \\ University of Pennsylvania, Philadelphia, PA 19104, USA }
\maketitle
\begin{abstract}
In this fluid dynamics video, we show that low Reynolds number swimmers, such as {\it Caenorhabditis (C.) elegans}, synchronize their gait when swimming in close proximity to maximize utilization of space. Synchronization most likely results from steric hindrance and enhances the propulsive speed only marginally.
\end{abstract}

% main text
\section{Main Text}
\verb'   'In this fluid dynamics video, we examine whether low Reynolds number swimmers such as {\it Caenorhabditis (C.) elegans} synchronize their gait when they are swimming in close proximity. The experimental apparatus consists of a tapered conduit. Two animals are introduced into the conduit, one after the other. The conduit is subjected to a DC electric field, with the negative pole at the narrow end and applied flow directed from the narrow end. As a result of their attraction to the negative pole (electrotaxis \cite{1}) of the electric field, both animals swim upstream towards the narrow end of the conduit. As the conduit narrows, the average adverse flow velocity increases and the swimming speed of the leading animal decreases faster than that of the trailing animal, allowing the trailing animal to catch up with the leading animal.\\

We quantify synchronization by measuring the phase lag between the gait of the trailing animal and the extended wave pattern of the leading animal as a function of the distance between the two animals. When the distance between the two animals is greater than one body length, the animals’ motions appear independent of each other (unsynchronized). Only when the distance between the two animals’ body centers is nearly equal to or smaller than one body length are the animals’ motions synchronized. When the nematodes are parallel to one another, synchronization enables the animals to swim together in a confined space without interference.\\
 
To examine whether synchronization is a result of steric hindrance, we carried out simple stochastic numerical experiments in which two animal-like objects with randomly-selected gaits were placed randomly in a conduit of dimensions similar to the one used in our experiments. We excluded events in which the two swimmers partially overlapped and calculated the synchronization as a function of the distance between the two swimmers. The model predictions nearly duplicate our experimental observations, lending credence to the notion that synchronization results from volume exclusion rather than a deliberate action of the animals.\\

Clearly, synchronization is essential to allow the two animals to reside in the same segment of the conduit, but would synchronization confer an advantage when one animal is trailing the other? To answer this question, we carried out direct numerical simulations and calculated the propulsive velocities of a single swimmer and of two synchronized and unsynchronized swimmers travelling in a single file with the trailer’s head just behind the leader’s tail. We found that the velocities of the two swimmers, when one chases the other, are only marginally higher compared to that of a single swimmer.\\

In summary, we have shown that swimmers synchronize their gait to maximize utilization of space and that synchronization most likely results from steric hindrance.  Numerical simulations indicate that synchronization enhance the propulsive speed only marginally.

\end{document}